\documentclass[prl,superscriptaddress,showpacs,twocolumn]{revtex4}

\usepackage{graphicx}
\usepackage[latin1]{inputenc}
\usepackage{amsmath}

\newcommand{\fig}[1]{Fig.~\ref{#1}}

\newcommand{\sij}[1]{\sum_{\langle ij#1\rangle}}
\newcommand{\bit}{\begin{itemize}}
\newcommand{\eit}{\end{itemize}}
\newcommand{\Na}{NaV$_2$O$_5$}

\newcommand{\etal}{{\em et al.}}

\begin{document}

%%%%%%%%%%%%%%%%%%%%%%%%%%%%%%%%%%%%%%%%%%%%%%%%%%%%%%%%%%%%%%%%%%%%%%%%%%%%%%%%%%%%%%%
\title{Creation and destruction of a spin gap in
 weakly coupled quarter-filled ladders}
%%%%%%%%%%%%%%%%%%%%%%%%%%%%%%%%%%%%%%%%%%%%%%%%%%%%%%%%%%%%%%%%%%%%%%%%%%%%%%%%%%%%%%%

\author{B. Edegger}

\affiliation{Institut für Theoretische Physik, Universität
Frankfurt, D-60438 Frankfurt, Germany}
\affiliation{Institut für
Theoretische Physik, Technische Universität Graz, A-8010 Graz,
Austria}

\author{H.G. Evertz}
\affiliation{Institut für Theoretische Physik,
Technische Universität Graz,
A-8010 Graz, Austria}

\author{R.M. Noack}
\affiliation{Fachbereich Physik,
Philipps Universität Marburg,
D-35032 Marburg, Germany}

\begin{abstract}
We investigate weakly coupled quarter-filled ladders with model
parameters relevant for NaV$_2$O$_5$ using density-matrix
renormalization group calculations on an
extended Hubbard model coupled to the lattice.
NaV$_2$O$_5$ exhibits super-antiferroelectric charge order
with a zigzag pattern on each ladder.
We show that this order
%, with a periodicity of four ladders,
causes a spin dimerization along the ladder and 
is accompanied by a spin gap of the same magnitude as that observed
experimentally. The spin gap is destroyed again at large charge
order due to a restructuring of the spins. An analysis of an
effective spin model predicts a re-creation of the gap by
inter-ladder singlets
when the charge order increases further.
\end{abstract} %-------------------------------------------------------------------


\maketitle

\begin{figure}[t]
  \centering
  \includegraphics*[width=0.37\textwidth]{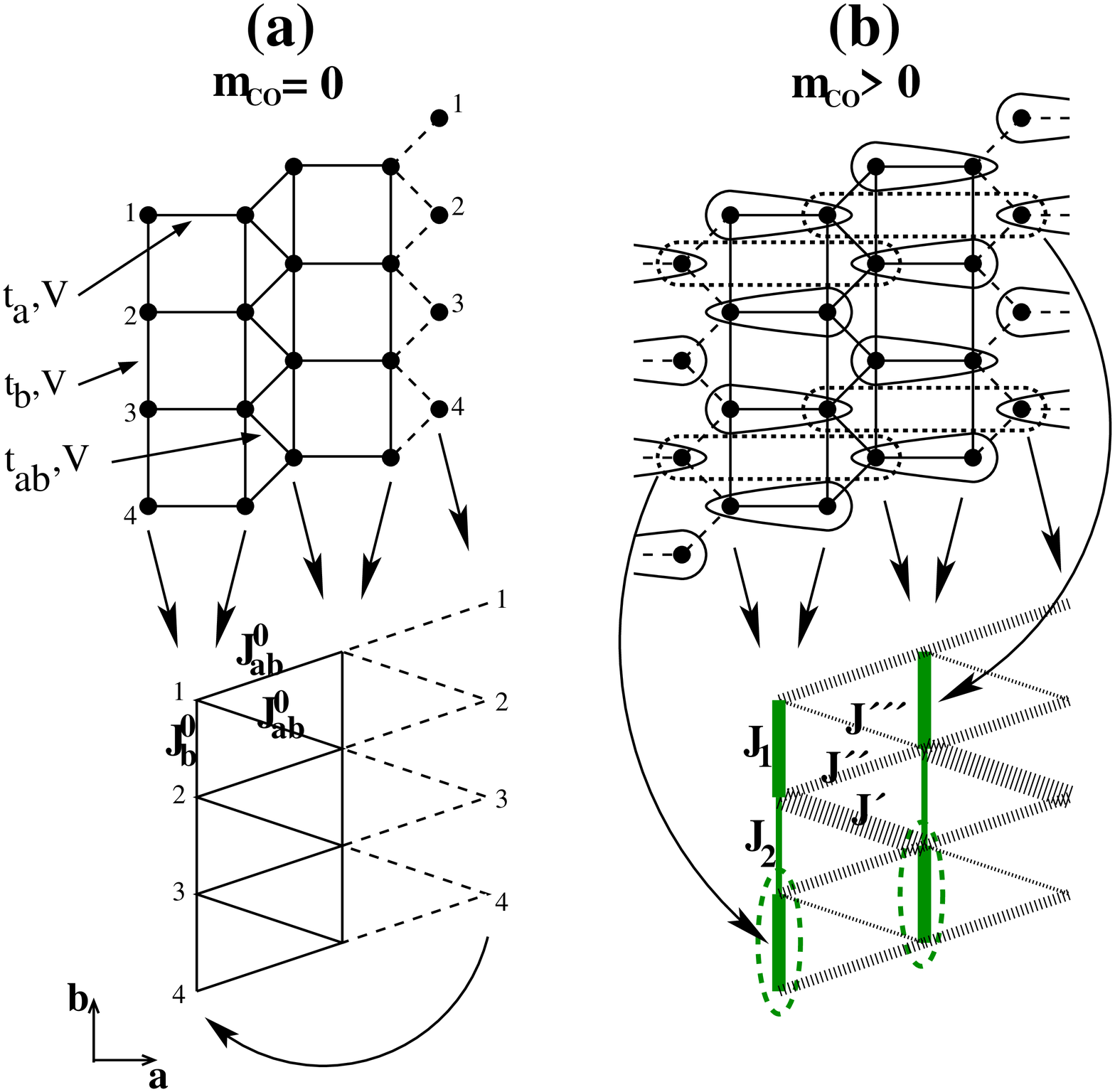}
  \caption{\label{fig:lattice}%
 Quarter-filled coupled ladders in \Na.
 The mapping of a rung (top) onto an effective spin site (bottom)
 is illustrated for (a) a charge-disordered and (b) a SAF-ordered regime.
 The shifted periodic boundary conditions in the $a$-direction are indicated by
 numbers which identify identical sites. The
 proposed singlet formation for \Na\ is depicted by dashed ellipses in
 the effective spin model (b, bottom).
}
\end{figure}

The discovery of a phase transition at $T_\text{C}\approx34K$ \cite{Isobe96} in \Na,
below which charge order and a spin gap appear \cite{Lemmens03}, has
precipitated
intensive theoretical investigation.
\Na\ consists of well-separated planes which contain weakly coupled
quarter-filled
vanadium ladders \cite{Smolinski98} (\fig{fig:lattice}).
They can be described by an
extended Hubbard model (EHM) \cite{Lemmens03}.
A zigzag charge order as observed in \Na\ \cite{Grenier01}
is already created in this model by the nearest-neighbor Coulomb repulsion $V$
for an isolated ladder, but only at overly large values of $V$
\cite{Seo99,Sa00,Vojta01,Aichhorn04,Edegger05,Gros05}.
In a recent DMRG study \cite{Edegger05}, we showed that the inclusion
of a strong
effective Holstein coupling to the lattice,
which was found in LDA calculations \cite{Spitaler04},
reduces the required Coulomb repulsion to a realistic value.
The DMRG calculations then yielded
good agreement between theoretical and experimental results for
the amount of charge order,  the extent of lattice distortion in
the $c$-direction, the effective spin coupling $J_\text{eff}$ in
the $b$-direction, and the charge gap.

However, on an isolated ladder \cite{Vojta01,Edegger05} the spin gap
vanishes in the thermodynamic limit.
The occurrence of the spin gap in \Na\ appears to be intimately connected
to the coupling of ladders,
as indicated, e.g., by the splitting
of magnon branches \cite{Yosihama98,Gros99,Grenier01}.
An intriguing scenario has been put
forward by Mostovoy and Khomskii \cite{Mostovoy00}.
It is based on the fact that the experimentally observed unit cell of
\Na\ in the ordered phase is
$2a\times 2b\times 4c$ \cite{Unitcell224}.
The charge order then has a periodicity of {\em four} parallel
ladders in the $a$-direction, and is thus super-antiferroelectric
(SAF) \cite{Chitov04}. The corresponding polarization of the electrons on the
rungs is illustrated by wide-narrow ellipses
in \fig{fig:lattice}(b) (top). Each ladder exhibits
antiferroelectric charge order along the ladder (generally called
zigzag charge order) as well as an antiferroelectric order to
the next nearest ladder. The ladders correspond to effective 
antiferromagnetic spin-1/2 chains in the $b$-direction.
For each ladder, the two
closest sites on the neighboring in-plane ladders alternate between
low charge occupation [indicated by dashed loops in
\fig{fig:lattice}(b)] and large occupation. Large charge
occupation should effectively weaken the electron hopping along
the ladder, both through purely electronic interactions as well
as by pushing away the neighboring oxygen atoms and thus reduce
the spin coupling.
With SAF charge order, the effective spin chains are therefore dimerized,
which was proposed to lead to the formation of spin singlets
[\fig{fig:lattice}(b), bottom]
and to the observed spin gap.
This scenario is difficult to evaluate quantitatively. It has been
investigated using exact diagonalization
\cite{Riera99,Grenier01,Gros05} on small clusters
and by means of an approximate $xy$-model together with a mean-field approximation \cite{Chitov04}.
It has also been argued that the mechanism proposed by Mostovoy
and Khomskii would not work for \Na\ \cite{Clay03}.

In the present paper, we use the Density Matrix Renormalization
Group (DMRG) to study the extended Hubbard model with coupling to
the lattice \cite{Edegger05} for large systems of coupled ladders
up to length 20, with periodicity of {\em four} ladders in the
$a$-direction. We show that the scenario proposed by Mostovoy and
Khomskii indeed works. In the thermodynamic limit, it produces a
dimerization and a spin gap of similar size to that seen in the
experiment. The spin gap 
is found {\em only} in the case of the four-ladder SAF order; it
does not appear for charge ordering with one- or two-ladder
periodicity as studied in previous papers
\cite{Edegger05,Aichhorn04}. Surprisingly, the spin gap closes
again at large Coulomb repulsion $V$. We explain this behavior by
considering an effective spin model for which DMRG calculations on
larger systems
%with lengths of up to 120 rungs
are possible.

{\em Model.} We use the extended Hubbard model with coupling to
the lattice as introduced in Ref.~\onlinecite{Aichhorn04}, $
H=H_\text{EHM}+H_\text{l}+H_\text{e-l} $, where
%\begin{align}
$ H_\text{EHM} = -\sum_{\langle ij\rangle,\sigma}t_{ij}
    \left(c_{i\sigma}^\dagger c_{j\sigma}^{\phantom{\dagger}}+\mbox{h.c.}\right)
%
%\label{hehm}
%\nonumber\\
%
    + U\sum_i n_{i\uparrow}n_{i\downarrow}+\sij{} V_{ij} n_i n_j~,$
%\end{align}
with hopping matrix elements
from first-principles calculations \cite{Spitaler04,Smolinski98,Horsch98},
$t_\text{a}=0.35\,eV,\,t_\text{b}=0.5 \,t_\text{a}$,
$t_\text{ab}=0.17\,t_\text{a}$ and $0.33\,t_\text{a}$, and a
uniform on-site repulsion $U=8.0\, t_\text{a}$. The lattice and
the corresponding model parameters are illustrated in
\fig{fig:lattice}(a). Since the nearest-neighbor Coulomb repulsion
$V$ is difficult to estimate, we investigate the functional
dependence on this parameter. Recent LDA calculations
\cite{Spitaler04} have shown that there is a very large
Holstein-like coupling to out-of-plane movements of oxygen atoms,
which can be modeled by including the terms
$H_\text{l}= \kappa\sum_i \frac{z_i^2}{2}$ %,\label{eq_2b}\\
and
$H_\text{e-l} = -C\sum_i z_i n_i$, % \; , \label{eq_2c}
with $\kappa=0.125 \,t_\text{a}$ and $C=0.35 \,t_\text{a}$, and
$z_i$ in units of $0.05$ \AA{}, in the Hamiltonian. We treat these
movements adiabatically like in Refs.~\onlinecite{Aichhorn04}
and~\onlinecite{Edegger05}. The optimal configuration for the
distortions $z_i$ is then a zigzag-pattern \cite{Edegger05} with
amplitude $z_\text{opt}$, for which we take the results from an
isolated ladder \cite{justify}. Then, $H_\text{l}$ becomes an
irrelevant constant and $H_\text{e-l}$ reduces to a zigzag
alternation of the local chemical potentials.

In order to investigate the proposal by Mostovoy and Khomskii, a
system with a periodicity of four ladders is necessary. Within the
DMRG, calculations on an extended Hubbard model of four coupled
ladders would be,
however, too restricted in length.
We therefore employ a system of
only two coupled ladders of length up to 20, but with {\em shifted
periodic boundary conditions}, as illustrated in \fig{fig:lattice}.
They ensure the proper periodicity and the same SAF structure of
neighboring charges in the ordered phase as in a four-ladder
system. We apply open boundary conditions\ (\emph{obc}) in the
$b$-direction.

%---------------------------------------------------------------------
\begin{figure}[t]
  \centering
  \includegraphics[width=0.39\textwidth]{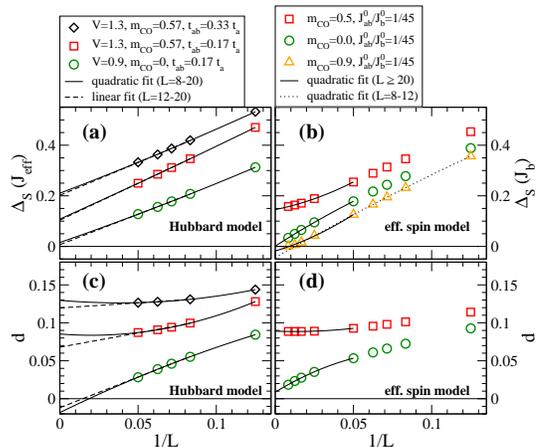}
  \caption{\label{fig:fss}%
Examples of finite-size extrapolations for (a,b) the spin gap $\Delta_\text{S}$
and (c,d) the dimerization $d$
of (a,c) the Hubbard model and of (b,d) the effective spin model.
Linear and quadratic fits in $\frac{1}{L}$ are illustrated
for charge-disordered  states ($m_\text{CO}=0$, circles)  and
SAF-ordered states ($m_\text{CO}>0$).
}
\end{figure}
%---------------------------------------------------------------------

{\em Results.} We calculate the spin gap
using \cite{Vojta01,error_truncation}
$\Delta_\text{S}(L) ~=~ E_0(L,N,S_z=1)-E_0(L,N,S_z=0)\, ,$
where $E_0$ is the ground state energy, $L$ is the length of each
ladder, $N=L/2$ is the number of electrons, and $S^z$ is the total spin
in $z$-direction. The results are extrapolated using linear and
quadratic fits in $1/L$, including $L=(8),12,14,16,20$, as illustrated
in \fig{fig:fss}(a). Due to broken translational invariance in the
$b$-direction (\emph{obc}), we can define a spin dimerization $d$
as
$d=\frac 1 2 \, \frac 1 4 \, \sum_{l={\rm I,II}} \sum_{i} \, (-1)^i \,
 \langle \hat S^{z}_{i,l} \,(\hat S^z_{i+1,l}-\hat S^z_{i-1,l})\rangle \ , $
where $i$ counts the rungs in the $b$-direction. We restrict
$\sum_{i}$ to four rungs in the middle of each ladder ($l={\rm
I,II}$). Here $S^z_{i,l}$ is the sum of the $S^z$-spin at the $i$-th
rung.
The dimerization $d$ is an indicator for bond alternation and
spin singlets along the ladder,
as illustrated in \fig{fig:lattice}(b) (bottom).
Translational invariance in the $b$-direction is additionally broken by the lattice
distortions, and we then find the contributions to $d$ to be positive on both ladders.
A state consisting of consecutive singlets in the $a$-direction
[\fig{fig:lattice}(b), bottom]
would give the maximum value for $d$, i.e., $|d_\text{max}|=1/4$.
%---------------------------
The charge order parameter $m_\text{CO}$ is given by \cite{Aichhorn04}
$m_{\text CO}^2=\frac{1}{N^2\langle n\rangle^2}\sum_{ij}e^{i{\mathbf
    Q}({\mathbf R}_i-{\mathbf R}_j)}\left(\langle n_i n_j\rangle -
       \langle n\rangle^2 \right) \, ,$
where ${\mathbf Q}=(\pi,\pi)$ and $N$ is the total number of sites
on the ladder.  We calculate $m_{\text CO}^2$ for each ladder
separately and then average.
%
%
%-----------------------------------------------------------------------------------
\begin{figure}
  \centering
  \includegraphics[height=0.4\textwidth
                   ]{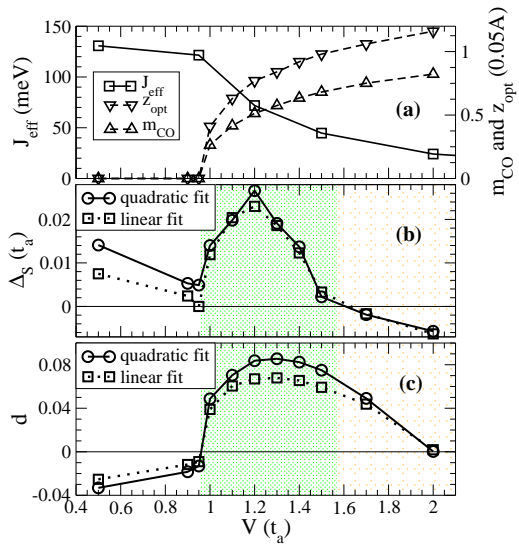}
  \caption{\label{fig:hubbard_prop}%
Results for the EHM with coupling to the lattice, as a function of $V$.
Extrapolated values of (a) the charge order parameter $m_\text{CO}$,
(b) the spin gap $\Delta_\text{S}$, and (c)
the dimerization $d$ are illustrated for two coupled ladders
($t_\text{ab}=0.17\,t_\text{a}$).
The effective magnetic exchange constant $J_\text{eff}$ \cite{Edegger05}
and the distortion $z_\text{opt}$ \cite{Edegger05} of an isolated ladder
are also shown in (a).
The three regimes discussed in the text
are highlighted.%
}
\end{figure}
%-----------------------------------------------------------------------------------
%
In the thermodynamic limit, we obtain the results shown in
\fig{fig:hubbard_prop} for the charge order parameter
$m_\text{CO}$, the spin gap $\Delta_\text{S}$, and the
dimerization $d$, as a function of $V$.
For comparison, we also show the lattice distortion $z_\text{opt}$
and the effective spin coupling $J_\text{eff}$, which were  determined for an isolated ladder
in Ref.~\onlinecite{Edegger05}.
We can identify three different regimes:
\\\emph{(i)} For small nearest-neighbor Coulomb repulsion
${V \le 0.95 \, t_\text{a}}$,
the lattice distortion and the charge order are both zero.
This regime is therefore not relevant for \Na.
The extrapolated values of the spin gap $\Delta_\text{S}$ are mostly finite,
but vary strongly between linear and quadratic extrapolations,
reflecting the difficulty of a reliable estimate in this regime.
However, in the 2d thermodynamic limit
we can expect a vanishing gap as long as $m_\text{CO}=0$,
from calculations on a system with pure \emph{pbc} in the
$a$-direction \cite{Edegger05} and from considerations of an effective
spin model (see below).
The dimerization $d$, which is expected to be zero in this regime, shows
slightly negative values, probably due to finite-size effects.
\\\emph{(ii)} At a critical $V_\text{C}\approx 0.95 \, t_\text{a}$
\cite{Edegger05},
the lattice distortions become finite and cause charge ordering in
a SAF pattern for our choice of boundary conditions. Linear and
quadratic extrapolations to $L=\infty$ match quite well and we find
that the spin gap $\Delta_\text{S}$ and the intra-ladder
dimerization $d$ are finite.
This is in contrast to the non-SAF charge-ordered system of two
coupled ladders with purely periodic boundary conditions studied
in Ref.~\onlinecite{Edegger05}, which shows \emph{neither} a
finite spin gap \emph{nor} finite dimerization in the
thermodynamic limit.
The concurrent appearance of intra-ladder dimerization and a spin
gap only when SAF charge ordering occurs strongly indicates that here charge
ordering indeed causes singlet formation along the ladder.
\\\emph{(iii)} At larger $V\approx 1.6 \, t_\text{a}$, the spin gap vanishes again,
to our initial surprise \cite{logarithm}. Concurrently, the dimerization becomes smaller,
but does not completely disappear.
This behavior and the negative extrapolations for the spin gap
are consistent with the effective spin system (see below).

Regime (\emph{ii}) can be associated with NaV$_2$O$_5$ because it
provides a description for the concurrent appearance of lattice
distortions, zigzag charge order, and the opening of a spin gap
in the low temperature phase.
At the
effective $V=1.3\,t_\text{a}$ previously determined \cite{Edegger05},
lattice
distortions, charge order, as well as the spin gap ($\Delta^{\rm exp}_\text{S}\approx
10\, meV = 0.029\,t_\text{a}$ \cite{Lemmens03})
are of the same magnitude as in \Na\ \cite{enhanceSG}.
Calculations for different inter-ladder hopping strengths $t_\text{ab}$
show that the maximal size of the gap increases with $t_\text{ab}$
($t_\text{ab}=0/0.17/0.33 \, t_\text{a} \Rightarrow
\Delta_\text{S,max} \approx 0/0.025/0.05 \, t_\text{a} $), but
that concurrently the transition from regime (\emph{ii}) to
(\emph{iii}) appears at smaller $V$ (for $t_\text{ab}=0.33 \,
t_\text{a}$ at $V\approx 1.4 \, t_\text{a}$), corresponding to
smaller $m_\text{CO}$.

\emph{Effective spin model.} In order to better understand the
behavior of the spin gap in the EHM,
we consider an effective spin model for NaV$_2$O$_5$ which was
introduced in Ref.~\onlinecite{Gros99}.
This model allows us to investigate much larger systems -- up to
length $L=120$ --
and thus to extrapolate results to the thermodynamic limit more accurately.
By replacing the quarter-filled rungs (one electron on each rung)
by a single effective spin site, we obtain the Heisenberg spin
model whose lattice structure is depicted in
\fig{fig:lattice}(a) (bottom) with two different magnetic
exchange constants, $J^0_\text{b}$ along the effective chains, and
$J^0_\text{ab}$, inter-chain. The chains in this effective spin
model correspond to the ladders of the original system.
In the SAF charge-ordered state with weakly coupled effective chains
[\fig{fig:lattice}(b)], the magnetic exchange $J^0_\text{ab}$
differentiates to three different values depending on the positions of the
electrons on the interacting rungs \cite{Gros99}: large
$J'=J^0_\text{ab}\,(1+m_\text{CO})^2$, medium
$J''=J^0_\text{ab}\,(1-m_\text{CO}^2)$, and small
$J'''=J^0_\text{ab}\,(1-m_\text{CO})^2$, as illustrated in
\fig{fig:lattice}(b). The bond alternation along the $b$-direction
is taken to be $J_{1/2}=J_\text{b} \,(1 ± \delta)$, where
$J_\text{b}=J^0_\text{b}\,(1-m_\text{CO}^2)$, and we set
$\delta(m_\text{CO}=0)=0$ and $\delta(m_\text{CO} \approx 0.44)
\approx 0.034$ \cite{Gros99}.
Assuming a linear dependence yields $\delta(m_\text{CO})
\approx 0.076 \, m_\text{CO}$.
Together with $J^0_\text{ab}/J^0_\text{b}=1/45$ \cite{Gros99} (or
$J^0_\text{ab}/J^0_\text{b}=1/25$),
where $J^0_\text{ab}$ and $J^0_\text{b}$ are the exchange constants at
$m_\text{\rm CO}=0$,
the parameters of the effective spin model are completely determined
and the behavior can be investigated as a function of 
the charge order parameter, $m_\text{\rm CO}$.

\begin{figure}[t]
  \centering
  \includegraphics[width=0.36\textwidth]{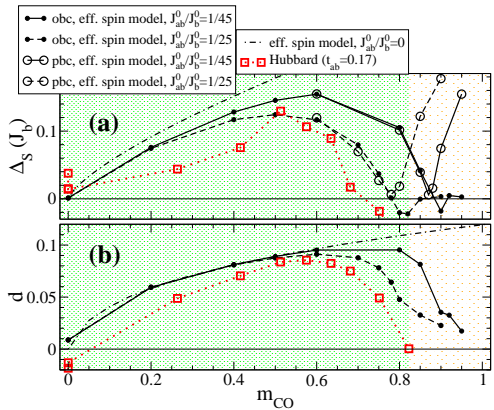}
  \caption{\label{fig:spin_sim}%
Effective spin model. Quadratically extrapolated values ($L=\infty$) for
(a) the spin gap $\Delta_\text{S}$ and
(b) the dimerization $d$,
as a function of the charge order parameter $m_\text{CO}$. Calculations with
\emph{obc} (black dots) and \emph{pbc} (open circles) in the
$b$-direction are shown.
For comparison the spin gap $\Delta_\text{S}$ for the EHM is shown,
in units of $J_\text{eff}$ [\fig{fig:hubbard_prop}(a)].}
\end{figure}

Like in the Hubbard system, we apply the DMRG to two coupled
spin chains (corresponding to two Hubbard ladders) with shifted
\emph{pbc}
in the a-direction, and \emph{obc} as well as
\emph{pbc} in the $b$-direction.
Due to the small inter-chain couplings, the finite-size effects
in the $a$-direction are small \cite{4CH}, justifying
calculations on only two coupled effective chains.
A quadratic extrapolation [\fig{fig:fss}(b),(d)] of the finite system
sizes ($L=20-120$ for \emph{obc}, $L=20-60$ for \emph{pbc}) to the
thermodynamic limit provides the $m_\text{CO}$-dependence of
dimerization $d$ and spin gap $\Delta_\text{S}$. The dimerization
and the spin gap increase with $m_\text{CO}$ up to a maximum
(\fig{fig:spin_sim}); then, both decrease
even though the bond alternation still increases.
This behavior
matches well with the results from the EHM. In contrast, uncoupled
chains (1D spin-1/2-Heisenberg chains) show a continuing increase
of dimerization and spin gap.

The destruction of the spin gap \cite{logarithm} is likely due to a
restructuring of the singlets,
which now occur
with a larger probability
on the $J'$-bonds.
For small charge ordering, all
inter-chain exchanges are of comparable size and are small in units of
$J_\text{b}$. The singlet formation along the $b$-direction is
then only marginally influenced. On the other hand, a large value
of $m_\text{CO}$ drastically changes the inter-chain
exchanges, e.g., for $m_\text{CO}=0.9$ and
$J^0_\text{ab}/J^0_\text{b}=1/45$, they become $J'/J_\text{b}
\approx 0.4$, $J''/J_\text{b} \approx 0.02$, and
$J'''/J_\text{b}\approx 0.001$. The large inter-chain
exchange $J'$ destroys the spin gap and causes a restructuring of
the singlets. For \emph{pbc} along the ladder [open circles in
\fig{fig:spin_sim}(a)] a re-creation of the spin gap is observed
after passing a critical charge ordering $m^*_\text{CO}$. In
contrast, \emph{obc}-calculations (black dots) show a vanishing
gap for all $m_\text{CO} \ge m^*_\text{CO}$, although both
calculations match excellently for $m_\text{CO}<m^*_\text{CO}$
[\fig{fig:spin_sim}(a)]. This strongly suggests a singlet formation
on the $J'$-bonds \cite{Seo99} with increasing $J'$.
In the case of \emph{obc}, unpaired spins are then present
at the ends of the ladders [\fig{fig:lattice}(b)], preventing a spin gap.
This explanation is supported by a
considerable increase of inter-chain spin correlations on
the $J'$-bonds for $m_\text{CO}>m^*_\text{CO}$.

For \emph{obc}, after passing the critical charge order
$m^*_\text{CO}$, the spin gap extrapolates to values that are
slightly negative [\fig{fig:spin_sim}(a)]. If the extrapolation is
restricted to smaller system sizes [dotted line in
\fig{fig:fss}(b)], the values become more negative. This is in
agreement with the negative extrapolations for $\Delta_\text{S}$
in the Hubbard model in the large $m_\text{CO}$-regime
[\fig{fig:hubbard_prop}(b)]. The magnitude of these negative values
gives a minimal estimate of the error in the extrapolation.

{\em Conclusions.} We have studied coupled quarter-filled ladders
in an extended Hubbard model with lattice coupling and with model
parameters appropriate for NaV$_2$O$_5$. Our calculations match
the effective spin model proposed by Gros and
Valenti \cite{Gros99}, and show two ways to realize a spin gap
for coupled quarter-filled ladder compounds.
At lower charge order we find a super-antiferroelectric 
regime with spin dimerization along the ladder and a spin gap,
in agreement with the scenario of Mostovoy and Khomskii \cite{Mostovoy00}.
This regime provides the possibility to understand the concurrent
appearance of charge order, lattice distortions, and spin gap in
NaV$_2$O$_5$ and is in good agreement with experimental
observations for \Na.

Higher charge order, unrealistically large for NaV$_2$O$_5$,
results in an increased inter-ladder magnetic exchange
and a restructuring of the spins. The  corresponding destruction
and re-creation of the spin gap illustrates an interesting
transition between different spin gap regimes. While regime
\emph{(ii)} is well described by weakly coupled rectangular
ladders with dimerization inducing SAF charge order, regime
\emph{(iii)} displays weakly coupled zigzag ladders with a charge
bond density wave \cite{Clay05charge-ordered}.

%{\em Acknowledgements:}
 This work was supported by the Austrian science fund FWF, project P15520.
 We thank M.\ Aichhorn, C.\ Gros, T.C.\ Lang, and F.\ Michel for
 useful conversations.
%

%%%%%%%%%%%%%%%%%%%%%%%%%%%%%%%%%%%%%%%%%%%%%%%%%%%%%%%%%%%%


\begin{thebibliography}{99}
\bibitem{error_truncation}
Up to 800 states were kept in the DMRG calculations. 
We estimate the errors (except for extrapolated values) to be smaller
than the symbol sizes in all figures.
\bibitem{logarithm} The vanishing of the spin gap is also indicated by a sign change
                    in the logarithmic corrections to $1/L$ scaling, 
                    both in the Hubbard model and in the effective spin model.
\end{thebibliography}
\end{document}